\documentclass[twocolumn]{aastex631}

\usepackage{gensymb}

\begin{document}

\title{Life beyond 30: probing the $-20<M_\mathrm{UV}<-17$ luminosity function at $8<z<13$ with the NIRCam parallel field of the MIRI Deep Survey}

\author[0000-0003-4528-5639]{Pablo G. P\'erez-Gonz\'alez}
\affiliation{Centro de Astrobiolog\'{\i}a (CAB), CSIC-INTA, Ctra. de Ajalvir km 4, Torrej\'on de Ardoz, E-28850, Madrid, Spain}

\author[0000-0001-6820-0015]{Luca Costantin}
\affiliation{Centro de Astrobiolog\'{\i}a (CAB), CSIC-INTA, Ctra. de Ajalvir km 4, Torrej\'on de Ardoz, E-28850, Madrid, Spain}

\author[0000-0001-5710-8395]{Danial Langeroodi}
\affiliation{DARK, Niels Bohr Institute, University of Copenhagen, Jagtvej 128, 2200 Copenhagen, Denmark}

\author[0000-0002-5104-8245]{Pierluigi Rinaldi}
\affiliation{Kapteyn Astronomical Institute, University of Groningen, P.O. Box 800, 9700 AV Groningen, The Netherlands}

\author[0000-0002-8053-8040]{Marianna Annunziatella}
\affiliation{Centro de Astrobiolog\'{\i}a (CAB), CSIC-INTA, Ctra. de Ajalvir km 4, Torrej\'on de Ardoz, E-28850, Madrid, Spain}

\author[0000-0002-7303-4397]{Olivier Ilbert}
\affiliation{Aix Marseille Univ, CNRS, CNES, LAM, Marseille, France}

\author[0000-0002-9090-4227]{Luis Colina}
\affiliation{Centro de Astrobiolog\'{\i}a (CAB), CSIC-INTA, Ctra. de Ajalvir km 4, Torrej\'on de Ardoz, E-28850, Madrid, Spain}

\author[0000-0003-1693-2868]{Hans Ulrik N{\o}rgaard-Nielsen}
\affiliation{Cosmic Dawn Center (DAWN), Denmark}
\affiliation{DTU Space, Technical University of Denmark, Elektrovej, Building 328, 2800, Kgs. Lyngby, Denmark}

\author[0000-0002-2554-1837]{Thomas R. Greve}
\affiliation{Cosmic Dawn Center (DAWN), Denmark}
\affiliation{DTU Space, Technical University of Denmark, Elektrovej, Building 328, 2800, Kgs. Lyngby, Denmark}
\affiliation{Dept.~of Physics and Astronomy, University College London, Gower Street, London WC1E 6BT, United Kingdom}

\author[0000-0002-3005-1349]{G{\"o}ran {\"O}stlin}
\affiliation{Department of Astronomy, Stockholm University, Oscar Klein Centre, AlbaNova University Centre, 106 91 Stockholm, Sweden}

\author[0000-0001-7416-7936]{Gillian Wright}
\affiliation{UK Astronomy Technology Centre, Royal Observatory Edinburgh, Blackford Hill, Edinburgh EH9 3HJ, UK}

\author[0000-0001-6794-2519]{Almudena Alonso-Herrero}
\affiliation{Centro de Astrobiología (CAB), CSIC-INTA, Camino Bajo del Castillo s/n, E-28692 Villanueva de la Cañada, Madrid, Spain}

\author[0000-0002-7093-1877]{Javier Álvarez-Márquez}
\affiliation{Centro de Astrobiolog\'{\i}a (CAB), CSIC-INTA, Ctra. de Ajalvir km 4, Torrej\'on de Ardoz, E-28850, Madrid, Spain}

\author[0000-0001-8183-1460]{Karina I. Caputi}
\affiliation{Kapteyn Astronomical Institute, University of Groningen, P.O. Box 800, 9700 AV Groningen, The Netherlands}

\author[0000-0001-6049-3132]{Andreas Eckart}
\affiliation{I.Physikalisches Institut der Universit\"at zu K\"oln, Z\"ulpicher Str. 77, 50937 K\"oln, Germany}

\author[0000-0001-5891-2596]{Olivier Le Fèvre}
\affiliation{Aix Marseille Univ, CNRS, CNES, LAM, Marseille, France}

\author[0000-0002-0690-8824]{\'Alvaro Labiano}
\affiliation{Telespazio UK for the European Space Agency, ESAC, Camino Bajo del Castillo s/n, 28692 Villanueva de la Ca\~{n}ada, Spain}

\author[0000-0003-4801-0489]{Macarena García-Marín}
\affiliation{European Space Agency, Space Telescope Science Institute, Baltimore, Maryland, USA}

\author[0000-0002-4571-2306]{Jens Hjorth}
\affiliation{DARK, Niels Bohr Institute, University of Copenhagen, Jagtvej 128, 2200 Copenhagen, Denmark}

\author[0000-0002-7612-0469]{Sarah Kendrew}
\affiliation{European Space Agency, Space Telescope Science Institute, Baltimore, Maryland, USA}

\author[0000-0002-0932-4330]{John P. Pye}
\affiliation{School of Physics \& Astronomy, Space Research Centre, Space Park Leicester, University of Leicester, 92 Corporation Road, Leicester, LE4 5SP, UK}

\author{Tuomo Tikkanen}
\affiliation{School of Physics \& Astronomy, Space Research Centre, Space Park Leicester, University of Leicester, 92 Corporation Road, Leicester, LE4 5SP, UK}

\author[0000-0001-5434-5942]{Paul van der Werf}
\affiliation{Leiden Observatory, Leiden University, PO Box 9513, 2300 RA Leiden, The Netherlands}

\author[0000-0003-4793-7880]{Fabian Walter}
\affiliation{Max-Planck-Institut f\"ur Astronomie, K\"onigstuhl 17, 69117 Heidelberg, Germany}

\author[0000-0003-1810-0889]{Martin Ward}
\affiliation{Centre for Extragalactic Astronomy, Durham University, South Road, Durham DH1 3LE, UK}

\author[0000-0001-8068-0891]{Arjan Bik}
\affiliation{Department of Astronomy, Stockholm University, Oscar Klein Centre, AlbaNova University Centre, 106 91 Stockholm, Sweden}

\author[0000-0002-3952-8588]{Leindert Boogaard}
\affiliation{Max Planck Institut f\"ur Astronomie, K\"onigstuhl 17, D-69117, Heidelberg, Germany}

\author[0000-0001-8582-7012]{Sarah E.~I.~Bosman}
\affiliation{Max-Planck-Institut f\"ur Astronomie, K\"onigstuhl 17, 69117 Heidelberg, Germany}

\author[0000-0003-2119-277X]{Alejandro Crespo G\'omez}
\affiliation{Centro de Astrobiolog\'{\i}a (CAB), CSIC-INTA, Ctra. de Ajalvir km 4, Torrej\'on de Ardoz, E-28850, Madrid, Spain}

\author[0000-0001-9885-4589]{Steven Gillman}
\affiliation{Cosmic Dawn Center (DAWN), Denmark}
\affiliation{DTU Space, Technical University of Denmark, Elektrovej, Building 328, 2800, Kgs. Lyngby, Denmark}

\author[0000-0001-8386-3546]{Edoardo Iani}
\affiliation{Kapteyn Astronomical Institute, University of Groningen, P.O. Box 800, 9700 AV Groningen, The Netherlands}

\author[0000-0002-2624-1641]{Iris Jermann}
\affiliation{Cosmic Dawn Center (DAWN), Denmark}
\affiliation{DTU Space, Technical University of Denmark, Elektrovej, Building 328, 2800, Kgs. Lyngby, Denmark}

\author[0000-0003-0470-8754]{Jens Melinder}
\affiliation{Department of Astronomy, Stockholm University, Oscar Klein Centre, AlbaNova University Centre, 106 91 Stockholm, Sweden}

\author[0000-0001-5492-4522]{Romain A. Meyer}
\affiliation{Max Planck Institut f\"ur Astronomie, K\"onigstuhl 17, D-69117, Heidelberg, Germany}

\author[0000-0002-3305-9901]{Thibaud Moutard} 
\affiliation{Aix Marseille Univ, CNRS, CNES, LAM, Marseille, France}

\author[0000-0001-7591-1907]{Ewine van Dishoek}
\affiliation{Leiden Observatory, Leiden University, PO Box 9513, 2300 RA Leiden, The Netherlands}

\author[0000-0002-1493-300X]{Thomas Henning}
\affiliation{Max Planck Institut f\"ur Astronomie, K\"onigstuhl 17, D-69117, Heidelberg, Germany}

\author{Pierre-Olivier Lagage}
\affiliation{Université Paris-Saclay, Université Paris Cité, CEA, CNRS, AIM, 91191, Gif-sur-Yvette, France}

\author[0000-0001-9818-0588]{Manuel Guedel}
\affiliation{Dept. of Astrophysics, University of Vienna, Türkenschanzstr 17, A-1180 Vienna, Austria}
\affiliation{Max Planck Institut f\"ur Astronomie, K\"onigstuhl 17, D-69117, Heidelberg, Germany}
\affiliation{ETH Zürich, Institute for Particle Physics and Astrophysics, Wolfgang-Pauli-Str. 27, 8093 Zürich, Switzerland}

\author[0000-0002-9850-2708]{Florian Peissker}
\affiliation{I.Physikalisches Institut der Universit\"at zu K\"oln, Z\"ulpicher Str. 77, 50937 K\"oln, Germany}

\author[0000-0002-2110-1068]{Tom Ray}
\affiliation{School of Cosmic Physics, Dublin Institute for Advanced Studies, 31 Fitzwilliam Place, D02 XF86 Dublin, Ireland}

\author[0000-0002-1368-3109]{Bart Vandenbussche}
\affiliation{Institute of Astronomy, KU Leuven, Celestijnenlaan 200D, 3001 Leuven, Belgium}

\author[0000-0002-8365-5525]{\'Angela Garc\'ia-Argum\'anez}
\affiliation{Departamento de Física de la Tierra y Astrofísica, Facultad de CC Físicas, Universidad Complutense de Madrid, E-28040, Madrid, Spain}
\affiliation{Instituto de Física de Partículas y del Cosmos IPARCOS, Facultad de CC Físicas, Universidad Complutense de Madrid, 28040 Madrid, Spain}

\author[0000-0001-8115-5845]{Rosa Mar\'{\i}a M\'erida}
\affiliation{Centro de Astrobiolog\'{\i}a (CAB), CSIC-INTA, Ctra. de Ajalvir km 4, Torrej\'on de Ardoz, E-28850, Madrid, Spain}

\begin{abstract}
We present the ultraviolet luminosity function and an estimate of the cosmic star formation rate density at $8<z<13$ derived from deep NIRCam observations taken in parallel with the MIRI Deep Survey (MDS) of the Hubble Ultra Deep Field (HUDF), NIRCam covering the parallel field 2 (HUDF-P2). Our deep (40~hours) NIRCam observations reach a F277W magnitude of 30.8 ($5\sigma$), more than 2 magnitudes deeper than JWST public datasets already analyzed to find high redshift galaxies. We select a sample of 44 $z>8$ galaxy candidates based on their dropout nature in the F115W and/or F150W filters, a high probability for their photometric redshifts, estimated with three different codes, being at $z>8$, good fits based on $\chi^2$ calculations, and predominant solutions compared to $z<8$ alternatives. We find mild evolution in the luminosity function from $z\sim13$ to $z\sim8$, i.e., only a small increase in the average number density of $\sim$0.2~dex, while the faint-end slope and absolute magnitude of the knee remain approximately constant, with values $\alpha=-2.2\pm0.1$ and $M^*=-20.8\pm0.2$~mag. Comparing our results with the predictions of state-of-the-art galaxy evolution models, we find two main results: (1) a slower increase with time in the cosmic star formation rate density compared to a steeper rise predicted by models; (2) nearly a factor of 10 higher star formation activity concentrated in scales around 2~kpc in galaxies with stellar masses $\sim10^8$~M$_\odot$ during the first 350~Myr of the Universe, $z\sim12$, with models matching better the luminosity density observational estimations $\sim$150~Myr later, by $z\sim9$.
\end{abstract}


\section{Introduction}
\label{sec:intro}
Some of the oldest stars discovered in the Milky Way indicate a very early onset of star formation in the Universe \citep{2010ApJ...708.1398T}. Indeed, some inferred stellar ages are older than the Hubble time or very close \citep{2002ApJ...572..861C,2015A&A...575A..26C,2018ApJ...867...98S,2022JHEAp..36...27V}, within 100--200~Myr after (or even before!) the Big Bang. On a complementary approach, (the quite complex) stellar population modeling of stellar clusters and nearby galaxies also indicates very old formation ages for, e.g., the central regions or cores of some elliptical galaxies \citep{1995ARA&A..33..381F,2015MNRAS.448.3484M,2016ARA&A..54..597C}.

From the point of view of cosmological evolution, one of the most fundamental questions is how fast the Universe was able to start forming stars and galaxies. The {\em Hubble Space Telescope} ({\em HST}) was able to reach redshifts up to $z\sim11$ \citep{2016ApJ...819..129O,2021NatAs...5..256J}, just 400~Myr after the Big Bang, thus providing statistical samples of high-redshift galaxy candidates from which luminosity functions at $z=8$--10 could be built \citep[see review][and references therein, among them \citealt{2015ApJ...803...34B,2015ApJ...810...71F,2016MNRAS.459.3812M,2018ApJ...868..115Y,2018ApJ...854...73I,2019MNRAS.486.3805B,2020ApJ...891..146R,2020MNRAS.493.2059B,2022arXiv220512980B}]{2022ARA&A..60..121R}.

With the launch of {\em JWST} on Christmas Day 2021, and the start of scientific operations on July 2022, several thorough multi-filter searches for the highest redshift galaxies have been carried out with the NIRCam instrument, mainly with datasets integrating for less than 2 hours per band. Tens of high redshift galaxy candidates have been reported at $z=10$--15, with a few at $15<z<18$, and even a handful at $z\sim20$ \citep{2023MNRAS.518.6011D,2022arXiv220709434N,2022ApJ...940L..55F,2023ApJ...946L..13F,2022arXiv221206666C,2022arXiv221102607B,2022arXiv221204480R,2023MNRAS.518L..19R,2022arXiv220801612H,2023MNRAS.519..157W,2023ApJ...942L...9Y}. However, different teams analyzing the same observations have only agreed on a few objects \citep{2022arXiv221206683B}. The spectroscopic confirmation of $z>10$ candidates is starting to be carried out with {\em JWST}/NIRSpec, however it may require exposure times of tens or even hundreds of hours \citep[][or taking advantage of lensing clusters, \citealt{2022arXiv221015639R,2022arXiv221015699W}]{2022arXiv221204568C,2023arXiv230315431A,2023arXiv230405378A}, thus bringing into question the feasibility of obtaining robust samples composed of hundreds or even tens of $z>10$ galaxies, with more promising results at $8<z<10$ \citep{2023arXiv230106811I,2023arXiv230106696S,2023arXiv230107072T,2023arXiv230109482F}.

Much more important than the race to discover the highest redshift galaxy or the oldest galaxy ever formed is the exploration of the first hundred million years of galaxy evolution. Constraining the evolution of galaxies within this primordial epoch  will provide constraints to simulations which still have significant uncertainties in a broad range of fundamental areas such as (super) massive black hole early growth or primordial black hole abundance, the nature and interplay between dark and baryonic matter in the Dark Ages \citep{2014MNRAS.439..300L,2017MNRAS.472.4414D}, or the mechanism followed by star formation fueled by pristine or low metallicity gas \citep{2011ApJ...740...13Z}, affecting properties such as the Initial Mass Function \citep[IMF;][]{2010ARA&A..48..339B,2012MNRAS.422.2246M}, and the feedback processes in primitive galaxies \citep{2022ApJ...939L..31H}.

In this Letter, we address the topic of identifying and characterizing the galaxy population at $z>8$ using new ultra-deep NIRCam observations. We observed for a factor 7x longer than  previous JWST observations (e.g., in the SMACS0723, GLASS, or CEERS fields), reaching 30--31~mag $5\sigma$ limiting magnitudes and enabling us to probe the high-redshift Universe nearly a factor of 10 fainter than the  previously analyzed observations in JWST cycle 1 surveys. This allowed us to constrain, for the first time, the faint end of the ultraviolet (UV) luminosity function from $z=8$ to $z\sim13$. 

This Letter is organized as follows. Section~\ref{sec:data} presents our dataset, the NIRCam parallel of the MIRI Deep Survey. Section~\ref{sec:selection} describes our method to select galaxies at $z>8$. The sample is used to construct UV luminosity functions and derive the early evolution of the luminosity density in the Universe, presented in Section~\ref{sec:lf}. Section~\ref{sec:conclusions} summarizes our findings. 

Throughout this Letter, we assume a flat cosmology with $\mathrm{\Omega_M\, =\, 0.3,\, \Omega_{\Lambda}\, =\, 0.7}$, and a Hubble constant $\mathrm{H_0\, =\, 70\, km\,s^{-1} Mpc^{-1}}$. We use AB magnitudes \citep{1983ApJ...266..713O}. All stellar mass and star-formation rate (SFR) estimations assume a universal \citet{2003PASP..115..763C} IMF.

\section{Description of the data}
\label{sec:data}

\begin{deluxetable*}{lcclcc}
\tablehead{\colhead{Filter} & \colhead{Area} & \colhead{$5\sigma$ depth} & \colhead{Filter} & \colhead{Area} & \colhead{$5\sigma$ depth}\\
& [arcmin$^2$] & [mag] & & [arcmin$^2$] & [mag]}
\startdata
HST/$F435W$   & 4.2  & $29.2^{29.4}_{29.0}$ & HST/$F125W$   & 3.5  & $29.9^{30.1}_{28.9}$\\
HST/$F606W$   & 7.2  & $29.9^{30.2}_{29.6}$ & HST/$F140W$   & 2.2  & $27.6^{27.7}_{27.5}$\\
HST/$F775W$   & 7.0  & $29.5^{29.7}_{29.3}$ & HST/$F160W$   & 3.4  & $29.6^{29.9}_{28.1}$\\
HST/$F850LP$  & 7.2  & $29.3^{29.5}_{29.0}$ & JWST/$F115W$  & 7.3 & $30.4^{30.6}_{30.1}$\\
HST/$F814W$   & 7.3 & $30.3^{30.5}_{29.9}$ & JWST/$F150W$  & 7.3 & $30.2^{30.4}_{29.9}$\\
HST/$F105W$   & 4.0  & $29.7^{30.0}_{28.5}$ & JWST/$F277W$  & 7.3 & $30.8^{31.1}_{30.4}$\\
               &     &                     & JWST/$F356W$  & 7.3 & $30.8^{31.0}_{30.4}$\\
\enddata
\caption{\label{tab:data}Table with information for all filters used in this work. We show the area in common with the MIRI European Consortium Guaranteed Time Observations MIRI Deep Survey NIRCam parallel observations and the 5$\sigma$ depth corresponding to a point-like source measured in a 0.3\arcsec\, diameter circular aperture and corrected for the limited aperture using empirical PSFs. Median and quartiles are provided for the magnitude limits, illustrating the varying depth within the analyzed sky region.}
\end{deluxetable*}

This Letter is based on the analysis of the NIRCam data taken in parallel with the Guaranteed Time Observations (GTO) of the Hubble Ultra Deep Field (HUDF) carried out in December 2022 by the MIRI European Consortium Team, Program ID 1283 (PIs Hans Ulrik N{\o}rgaard-Nielsen R.I.P., and G{\"o}ran {\"O}stlin). While MIRI was observing the HUDF with the F560W filter during 61 hours (including overheads) in our MIRI Deep Survey (MDS, {\"O}stlin et al., in prep.), parallel data were also being acquired by the NIRISS (20 hours) and NIRCam (40 hours) instruments. 

The NIRCam observations were carried out with the F115W, F150W, F277W, and F356W filters. Nominal exposure times were 55187~s in each filter, achieved with 2 visits per filter using 20 integrations with NGROUP=7, DEEP8 readout pattern, and 10 dithering positions following the MIRI-optimized CYCLING medium-size pattern with different starting points in each visit. Unfortunately, one of the 4 visits scheduled for our NIRCam observations could not be executed due to several JWST safe-mode events in December 2022, and one more visit had a significant difference in the roll angle with respect to the first two completed observations ($23\degree$ vs$.$ $31\degree$). This resulted in a smaller area in common between the four filters: 8.0~arcmin$^2$ instead of the 9.7~arcmin$^2$ expected for a full NIRCam field of view (a further area cut is described in Section~\ref{sec:photometry}). The aborted observations also resulted in shorter exposure times for the data used in this paper: nominal exposure times of 55187~s for F115W, and F277W and 27594~s for F150W and F356W. 

The typical 5$\sigma$ depth of these observations (measured in a 0.3\arcsec-diameter circular aperture and corrected for the encircled energy fraction, $\sim75\%$) are 30.3~mag and 30.8~mag in the short and long wavelength channels. This is 2--3 magnitudes deeper than other surveys such as CEERS, GLASS, or SMACS0723 (see \citealt{2022arXiv221206683B}).

The NIRCam data were calibrated with the \texttt{jwst} pipeline version 1.8.4, reference files in pmap version 1023. Apart from the standard pipeline stages (including snowball and wisp correction), we also applied a background homogenization algorithm prior to obtaining the final mosaics, including $1/f$-noise removal. The whole dataset was registered to the same World Coordinate System (WCS) reference frame using the Hubble Legacy Field (HLF) catalog in \citet{2019ApJS..244...16W}, based on Gaia DR1.2 (\citealt{2016A&A...595A...1G}, \citealt{2016A&A...595A...2G}). All images were drizzled to the same plate scale, namely, 0.03 arcsec/pixel. The FWHM of the point spread function (PSF) is 0.07\arcsec, 0.07\arcsec, 0.12\arcsec, and 0.14\arcsec\, for F115W, F150W, F277W, and F356W, respectively. All images were PSF-matched to the reddest filter.


The MDS NIRCam-parallel observations target a region known as the HUDF-P2 region, a field where the deepest HST/Advanced Camera for Surveys (ACS) F814W data exist, obtained simultaneously while observing the HUDF with WFC3. The HUDF-P2 field is located to the south-east (SE) of the HUDF, at the edge of the CANDELS footprint \citep{2011ApJS..197...35G,2011ApJS..197...36K}. A varying fraction of the NIRCam field of view is covered by {\em HST} ACS and WFC3 observations, all counting with heterogeneous depths. We used the Hubble Legacy Field v2 images for all filters used by ACS and WFC3 \citep{2019ApJS..244...16W}. Table~\ref{tab:data} presents depths and area coverage for all the datasets used in this paper. 

\newpage 

\section{Selection of $z\gtrsim8$ galaxy candidates}
\label{sec:selection}

A robust selection of high-redshift galaxy candidates is hampered by three different systematic effects. The first one is the photometry, since we are dealing with intrinsically faint sources which should be almost or completely undetected in some bands (e.g., those blue wards of the Lyman break or the Lyman-$\alpha$ jump), thus complicating the spectral energy distribution (SED) analysis. The comparison of results obtained by analyzing photometry in different apertures can potentially help to get rid of noise or contamination by neighboring sources in the determination of the photometric redshifts, as we will explain in Section~\ref{sec:photometry}. The second difficulty is the degeneracy in the analysis of the SED for the candidates, given that large red colors consistent with the Lyman-$\alpha$ break can also be mimicked by the Balmer or 4000~\AA\, breaks as well as by the presence of strong (i.e., high equivalent width) emission lines or dusty starbursts \citep{2022arXiv220801816Z,2022arXiv220802794N,2023arXiv230100017M,2022arXiv221100045P}. The third problem, directly related to the previous one, is the limited and/or biased set of templates that different programs estimating photometric redshifts use. This translates to methodological differences in how the SED degeneracy is addressed. In order to cope with the last two difficulties, we not only used several SEDs for a given source (measured in different apertures) but we also estimated the photometric redshifts with different codes and {\it a priori} assumptions (including the set of templates, but also how to treat low signal-to-noise data points). This is described in Section~\ref{sec:photoz}.

\subsection{Photometric catalog}
\label{sec:photometry}

We selected galaxies in the F277W filter (the deepest) and measured photometry in all bands following a methodology especially developed to deal with ultra-deep data and crowded fields.

First, we masked the four bright stars present in one of the modules and the very bright spikes coming from two other stars located to the East of our pointing (outside the field of view, but the bright diffraction spikes protrude our observations). This resulted in the loss of 9\% of the field of view. The final MDS-NIRCam-par survey area is 7.3~arcmin$^{2}$.

We then detected sources in several passes, starting from the most extended ones and progressing to smaller and fainter galaxies in a number of levels. The galaxies detected in a given level were masked completely down to the isophote corresponding to 5$\sigma$ of the sky ($\sim26.3$~mag~arcsec$^{-2}$ in F277W), building a first segmentation map of galaxy cores. Afterwards, the segmentation map for each galaxy was extended to the fainter outskirts by fitting isophotes constructed by dilation of the galaxy cores. The outskirts were fitted down to 1$\sigma$ of the sky (around 28.0~mag~arcsec$^{-2}$ in F277W) or the \citet{1980ApJS...43..305K} aperture, whichever was reached first. The Kron factor was varied as we were progressing to smaller/fainter galaxies, starting at 2.5 for the brightest galaxies and reaching 1.2 for the faintest sources (see \citealt{2023ApJ...946L..13F}). These numbers were calibrated with the sky surface brightness limit mentioned above. The method allows us to select faint galaxies (or globular clusters) in the outskirts of bright extended objects and improve the background determination in the whole field. Once all galaxies in a given level were fitted, the residual image was constructed by removing the core and the outskirts fitted with the isophotal analysis, and the new frame was fed to the next level of detection. We implemented a total of 10 levels for our final catalog in this dataset, starting from galaxies covering more than 10\,000 pixels (10~arcsec$^{2}$) in the first pass to the faintest objects with just 30 pixels ($\sim$0.03~arcsec$^{2}$).

Once the procedure was completed for the selection band, the photometry was measured in all other filters by fixing centers, shapes, and sizes. The final catalog is composed of 40\,526 sources, of which 33\,905 were detected at the 5$\sigma$ level or higher in F277W. The photometric catalog included color measurements in the \citet{1980ApJS...43..305K} aperture as well as fixed-diameter circular apertures with sizes 0.2\arcsec, 0.3\arcsec, 0.4\arcsec, and 0.5\arcsec. The SEDs corresponding to these fixed apertures were scaled to the Kron aperture to obtain total integrated magnitudes by applying a constant offset to all bands in a galaxy-by-galaxy basis. 

Photometric errors were estimated by measuring the background noise locally around each source with a procedure similar to that presented in \citet{2008ApJ...675..234P}, devised to take into account correlated noise. We remark that estimating realistic  uncertainties not affected by drizzling effects (i.e., not underestimated; see, e.g., \citealt{2003AJ....125.1107L}, \citealt{2007AJ....134.1103Q}, \citealt{2008ApJ...675..234P} or \citealt{2023ApJ...946L..13F}) and consistently between the SW and LW channels (which are drizzled in a different way given their distinct nominal pixel sizes, see below) is of utmost importance for our selection and analysis of dropout sources, especially given that the observations for the bluest channels just provide upper limits for the flux.

We built artificial apertures composed of randomly-selected non-contiguous pixels, adding up the number of pixels of the photometric aperture whose uncertainty we were trying to estimate. We also imposed that each chosen pixel should be more than three pixels away from any other pixel entering the analysis, this distance being a good approximation of the area that contributes to the flux of each pixel in the final mosaics for any band (after drizzling). With this additional procedure added to a random selection strategy, we ensured that all the pixels included in the noise calculation are independent, thus minimizing the effects of correlated noise. We remark that in the selection of high-redshift galaxy candidates, we analyzed five different SEDs for each source using the fluxes and noise calculations in the mentioned apertures.

We also calculated photometric errors using randomly distributed circular apertures (as in \citealt{2022arXiv220801612H}, for example). With this method, we find values of the rms of the sky that are 60\% and 3\% smaller for the long and short wavelength channels, respectively, compared to our method based on randomly-selected non-contiguous pixels. We interpret this difference as the effect of correlated noise, which biases the noise calculation in randomly distributed circular apertures since they include contiguous pixels whose signal comes from the same original pixel (i.e., with nominal size). Our random-selection method avoids contiguous pixels, so the effect of correlated noise should be minimized. We note that the F277W and F356W images are drizzled to half the nominal pixel size (65~mas/pixel), while the nominal pixel size for the F115W and F150W filters (32~mas/pixel) is very similar to the value used in our final mosaics (30~mas). Therefore, we should expect a larger effect of correlated noise in the rms calculations with circular apertures for the long wavelength bands, in agreement with the results of our test.

\subsection{Photometric redshift estimation}
\label{sec:photoz}

The SEDs for each source were fitted with three different codes and techniques, each providing a redshift probability distribution function (zPDF). 

Our fiducial photometric redshift estimation came from running the \texttt{eazy} code \citep{2008ApJ...686.1503B} using the default FSPS templates \citep{2010ascl.soft10043C}, plus a dusty galaxy template \citep{2013ApJS..206....8M}. We added the templates presented by \citet{2022arXiv221110035L} to optimize the analysis of high-redshift galaxies, which include emission-line galaxies with high equivalent widths (known to contaminate {\em JWST} high-redshift samples, see \citealt{2022arXiv220801816Z} and \citealt{2022arXiv220802794N}) and a variety of UV slopes. We used a flat prior in F277W magnitude and no template error, allowing the redshift to take values between $z=0$ and $z=20$. Data points with low ($<3$) signal-to-noise ratio (SNR) were treated in two different ways. We either used directly measured fluxes (even if negative) or we also changed those values to 5$\sigma$ upper limits and used them in a modified version of {\sc eazy}. The modification consisted of not allowing any fit to provide brighter fluxes (achieved by penalizing the $\chi^2$ calculation) than those limits and excluding the band in the $\chi^2$ calculation for templates providing lower fluxes (\citealt{2022arXiv221100045P}, M\'erida et al.\ 2023, submitted). 

We also used the \texttt{prospector} SED-fitting algorithm \citep{prospector} to further verify the estimated photometric redshifts of our candidates. 
We adopted a setup similar to that used in \cite{2022arXiv221202491L} for fitting the stellar populations of spectroscopically-confirmed $z\sim8$ galaxies. 
In brief, the free parameters include the total formed stellar mass, the stellar metallicity, the nebular metallicity, the nebular ionization parameter, the dust attenuation \citep[modelled with three free parameters, adopting the dust model of][]{Kriek_2013}, and the optical depth of intergalactic medium \citep[see][for details]{2022arXiv221202491L}.
Unlike \cite{2022arXiv221202491L}, where the redshift was fixed to the spectroscopically measured value, here we treated redshift as a free parameter with a flat prior in the range $z$\,=\,0\,--\,20.
The star formation history (SFH) was modelled non-parametrically in five temporal bins, with the last bin spanning 0--10 Myr in lookback time and the rest evenly spaced in $\log$(lookback time) up to $z = 35$.
We sampled the parameter space using the \texttt{dynesty} sampler \citep{2020MNRAS.493.3132S, 2022zndo...6609296K}, a dynamic nested sampler based on \cite{2019S&C....29..891H}.

Finally, we used the template-fitting code \texttt{LePhare} \citep{arnouts_measuring_2002,ilbert_accurate_2006} to help with the selection of our candidates. We adopted two configurations of this program. For the first one, we followed \citet{Rinaldi_22, Rinaldi_23}. Briefly, we made use of galaxy templates with the following set of SFHs: a standard exponentially declining ($SFR (t) \propto \exp^{-(t-t_{0})/\tau}$) and an instantaneous burst by adopting a simple stellar population (SSP) model ($SFR (t) \propto \delta(t)$). In particular, for the standard exponentially declining models we used the following e-folding timescales ($\tau$) in Gyr: 0.01, 0.1, 0.3, 1, 3, 5, 10, 15. We employed the stellar population synthesis (SPS) models of \citet{bruzual_stellar_2003} based on a  \citet{2003PASP..115..763C} IMF, considering two different values of the metallicity: solar metallicity and one-fifth of solar metallicity. To account for the effects of internal dust extinction, we convolved the model templates with a modified version of the reddening law of \citet{calzetti2000}, where we adopted the extrapolation provided by \citet{Leitherer_2002} at shorter wavelengths. For that purpose, we allowed the color excess $E(B-V)$ to range from 0 to 1.5~mag (with steps of 0.1), reaching $\rm A_{V} \simeq 6$~mag. A second configuration was explored, this time following the one described in  \citet{ilbert_evolution_2015} and \citet[][]{Kauffmann22}. Since our aim is to select rare sources at $z>8$, we explore a large range of dust-attenuation in order to identify and reject possible contaminants. We consider two dust attenuation curves as a free parameter \citep{calzetti2000,arnouts_encoding_2013} with $E(B-V)$ varying from $0$ to $2$~mag (reaching $A_V=8$~mag). In this execution, we also adopted the recipes of \citet{saito_synthetic_2020} to include the emission lines, allowing for an additional free rescaling by a factor two of all line fluxes. We note that \texttt{LePhare}'s fits are performed based on the fluxes and not the magnitudes, which do not require the need of introducing upper limits, considering that the flux uncertainties at low signal-to-noise are still meaningful \citep[following][]{laigle_cosmos2015_2016}. We rejected sources with $\chi^2>100$ which corresponds to an unreliable fit and sources which are best fitted by a brown dwarf stellar template. 

Given that we had two executions of \texttt{eazy} and two of \texttt{LePhare}, in order to avoid a bias towards any of these codes, we considered the highest redshift among the two executions for each code to select galaxies as explained in the next subsection.

Summarizing, we searched for high-redshift galaxy candidates using five different SEDs and three photometric redshift codes and techniques.

\subsection{Selection of candidates}

\begin{figure*}[ht!]
\centering
\includegraphics[clip, trim=1.6cm 5.0cm 2.0cm 1.5cm,scale=0.6]{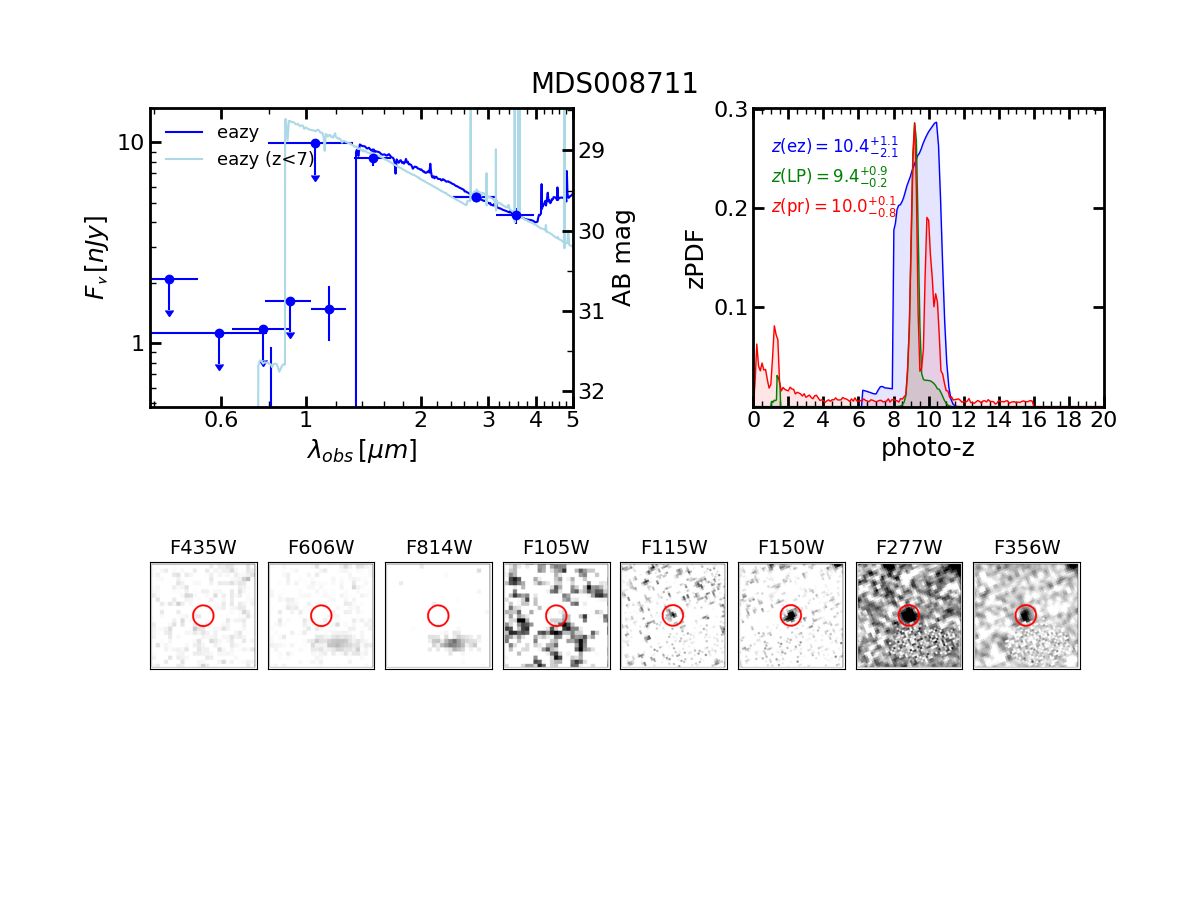}
\includegraphics[clip, trim=1.6cm 5.0cm 2.0cm 1.5cm,scale=0.6]{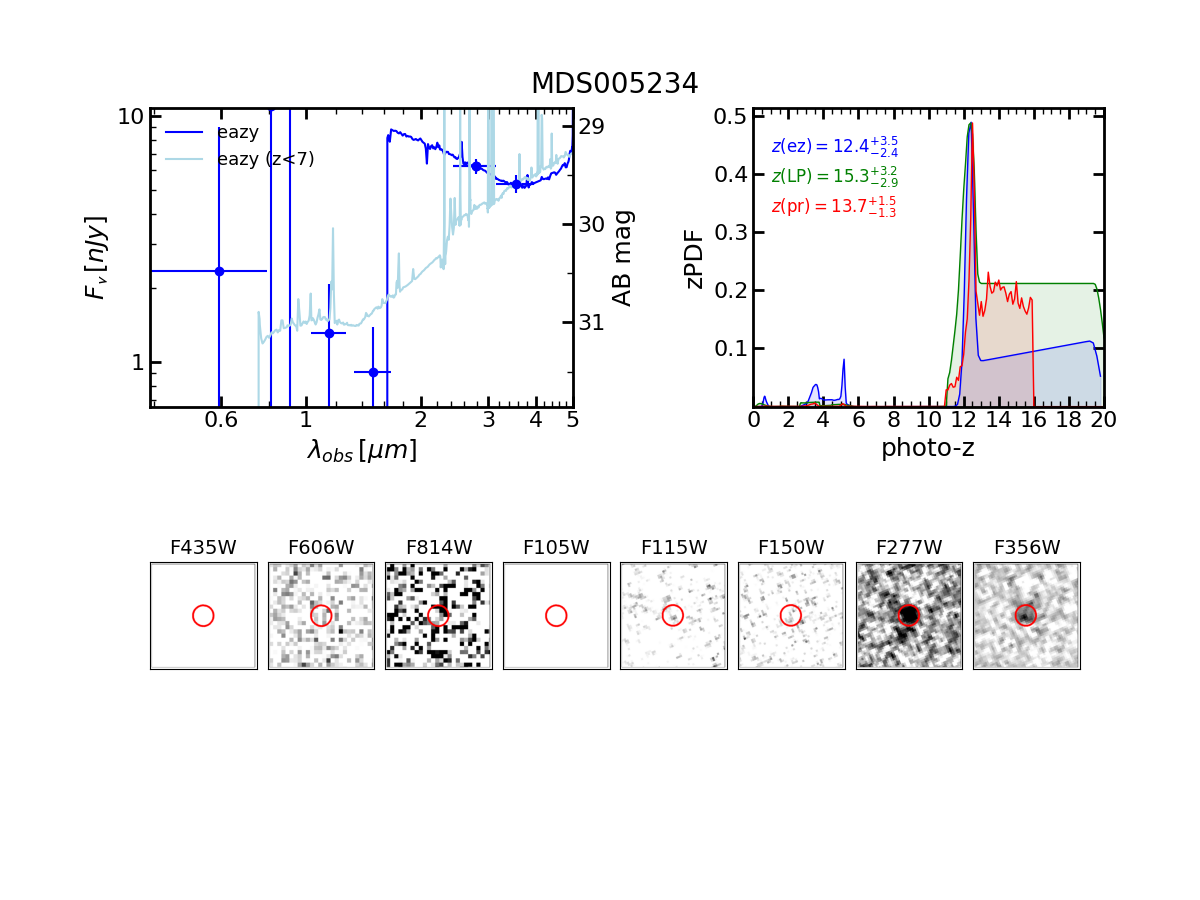}
\caption{\label{fig:stamps} Examples of F115W and F150W dropout galaxies in our sample. For each source, in the top left, we show the SED (for the 0.3\arcsec\, diameter aperture; filled dots) and template fits to low and high redshift solutions. Arrows indicate $1\sigma$ upper limits. The top right panel shows the photometric redshift probability distribution function for the different codes used in this work and the derived redshift and uncertainties. The bottom of the figure for each source presents $1.5\arcsec\times1.5\arcsec$ postage stamps in ACS, WFC3 and NIRCam bands, with the source marked with a 0.15\arcsec\, radius red circle.}
\end{figure*}

\begin{figure*}[ht!]
\centering
\includegraphics[clip, trim=1.6cm 5.0cm 2.0cm 1.5cm,scale=0.6]{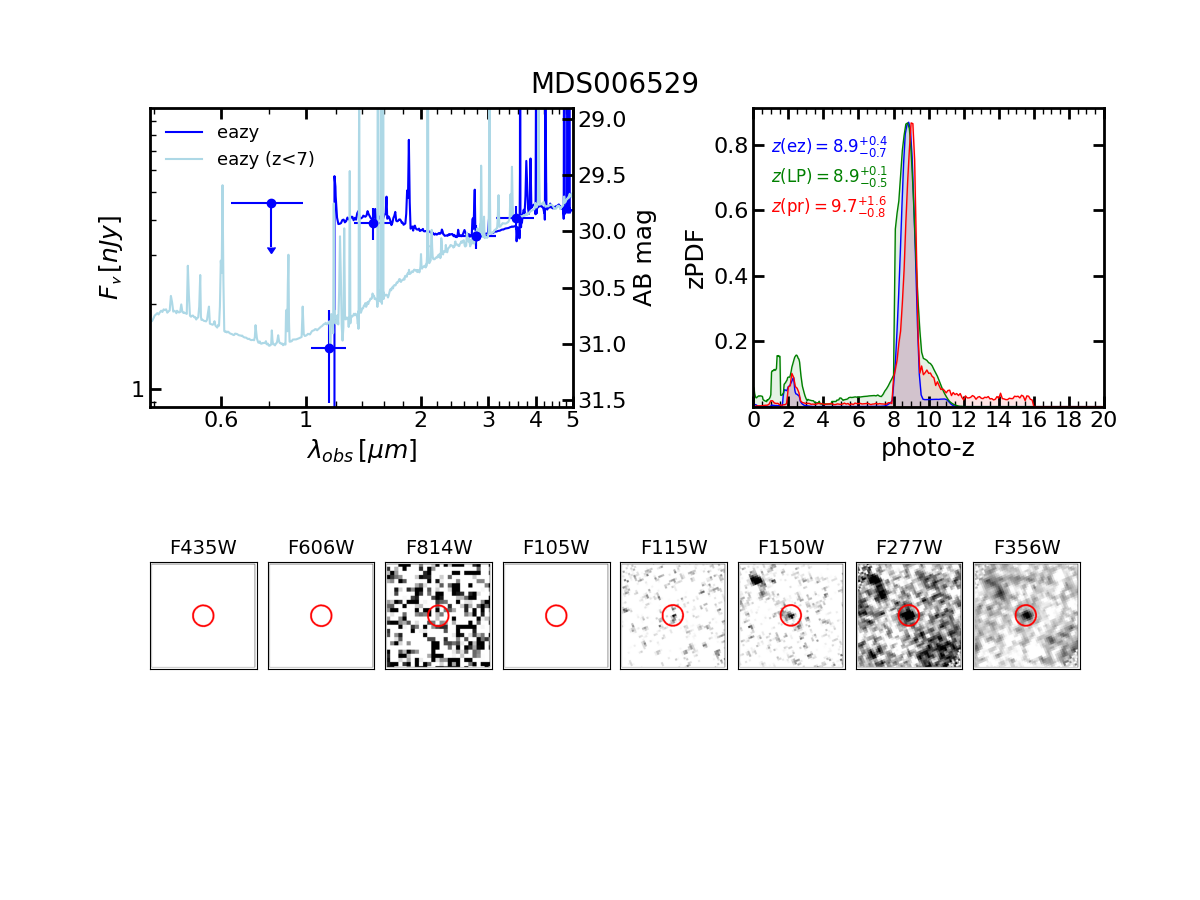}
\includegraphics[clip, trim=1.6cm 5.0cm 2.0cm 1.5cm,scale=0.6]{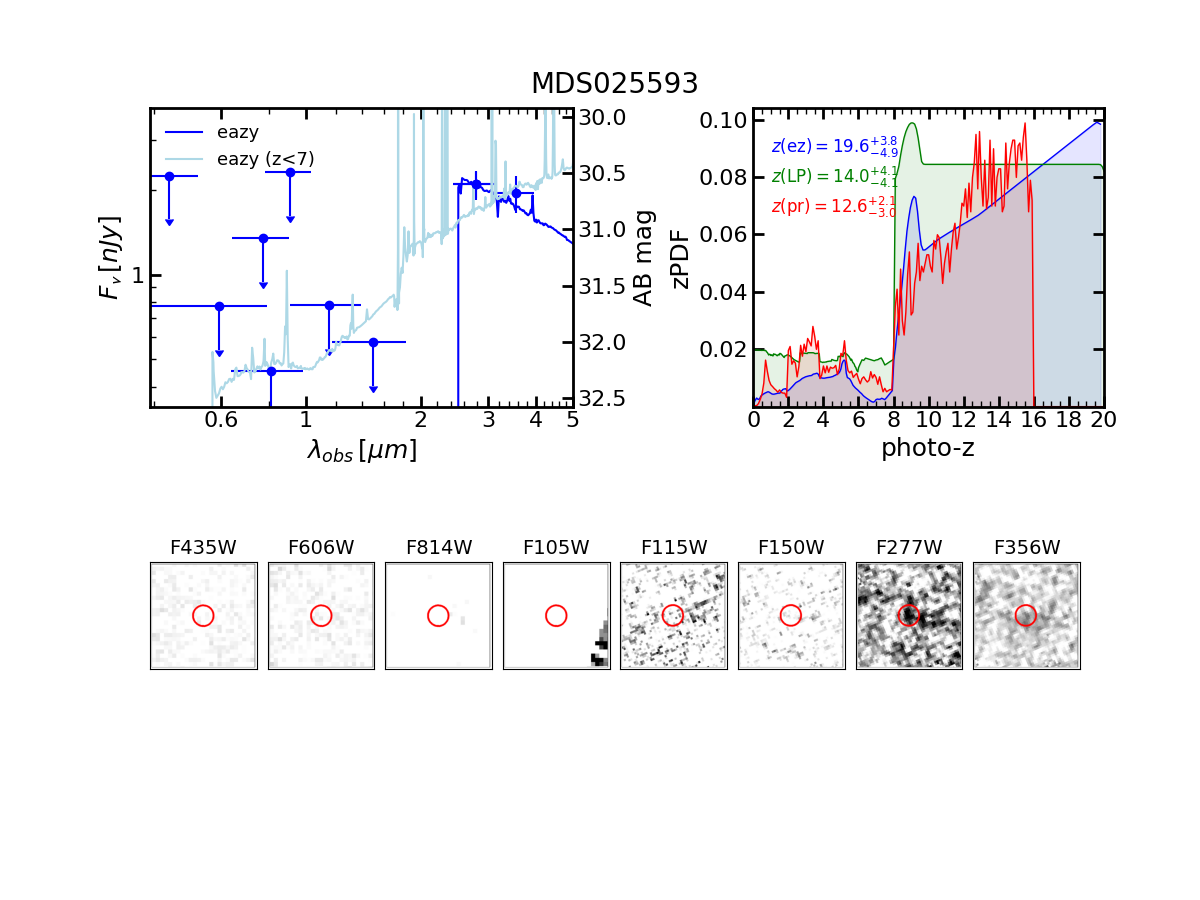}
\caption{\label{fig:stamps_faint} Examples of F115W and F150W dropout galaxies in our sample, in this case for sources fainter than $\mathrm{F277W}=30$~mag (distinctively probed by our dataset). Same information as in Figure~\ref{fig:stamps} given.}
\end{figure*}

We constructed a sample of $z\gtrsim8$ galaxy candidates by searching for F115W and F150W dropouts in our NIRCam data and applying a similar methodology to that presented in \citet{2023ApJ...946L..13F}. Distinctively, we explored the effect of the photometric aperture and the peculiarities of different photometric redshift estimation codes. Indeed, for each galaxy we used five SEDs derived from different aperture sizes and three redshift estimation codes, as outlined in the previous section, working with weighted averages for all relevant quantities used in the selection. We applied twice the weight of any other SED to that corresponding to 0.3\arcsec\, diameter aperture (typically used in this type of studies, see \citealt{2022arXiv221206683B}, \citealt{2022arXiv220801612H}, \citealt{2023MNRAS.518.6011D}, \citealt{2022arXiv220709434N} , \citealt{2023MNRAS.518.4755A}), which maximized the SNR of the measurements while keeping PSF-related aperture correction below 25\%. 

First, we selected all galaxies whose median F277W and F356W flux, averaged across all five SEDs, had a  $\mathrm{SNR}>5$. This cut resulted in a sample of 16\,133 sources. Then, we restricted the catalog to all sources with a median $\mathrm{SNR}<3$ in F115W and/or F150W (in both the PSF-matched and original images), resulting in a sample of 3883 1.15~$\mu$m dropouts and 2547 1.50~$\mu$m dropouts. 


\begin{figure*}[ht!]
\centering
\includegraphics[clip, trim=2.cm 3.0cm 7.0cm 1.9cm,scale=0.84]{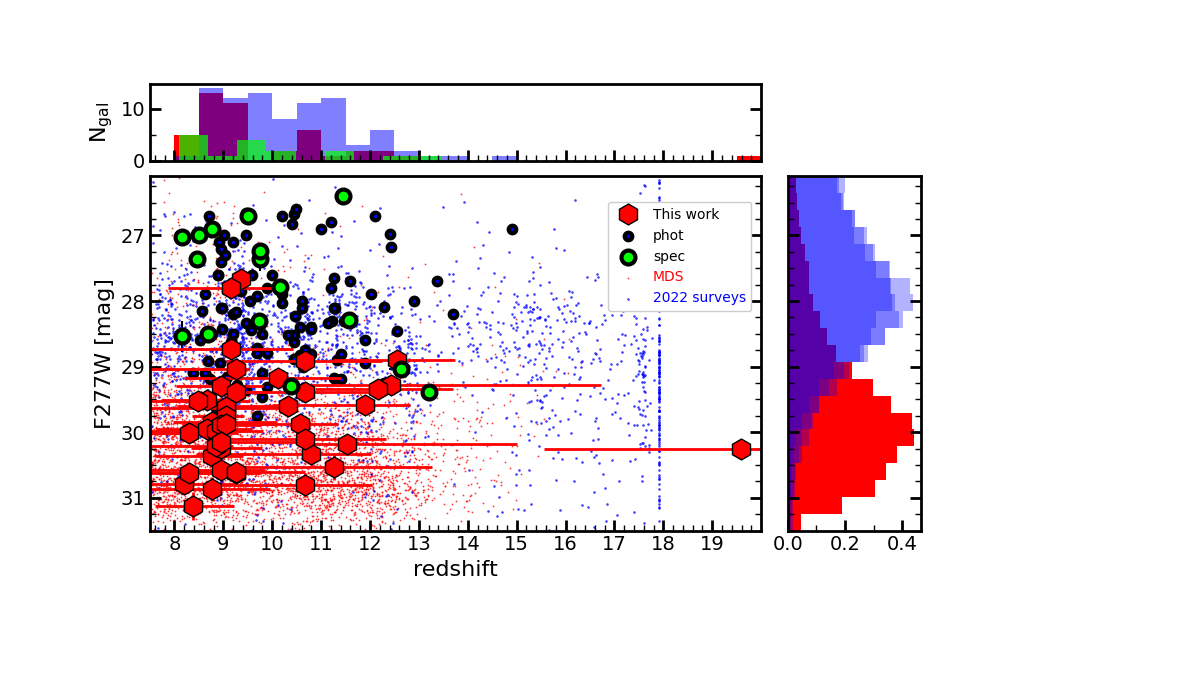}
\caption{\label{fig:z-mag}Magnitude in the NIRCam F277W filter vs$.$ photometric redshift for $z>8$ galaxies. General galaxy samples extracted from the catalogs released by G. Brammer for the SMACS0723, GLASS, and CEERS datasets gathered in 2022 are plotted with blue small dots  to illustrate the magnitude limits of the surveys and the number of high-redshift galaxies in unsupervised and unvetted photometric catalogs. Our parent galaxy sample is plotted with red small dots, and the final sample of $z>8$ galaxy candidates with red hexagons. We depict $1\sigma$ uncertainties plotted in both axes and the data point referring to the zPDF peak redshift. We also plot the photometric high-redshift galaxy candidates of \citet{2023MNRAS.518.6011D}, \citet{2022arXiv220709434N}, \citet{2022arXiv221206683B}, \citet{2022arXiv220801612H}, and \citet{2023ApJ...946L..13F} with large black circles (we do not remove repeated sources with different measurements), as well as spectroscopically confirmed galaxies from JADES \citep{2022arXiv221204568C,2022arXiv221204480R}, lensing clusters \citep[][not corrected for magnification]{2023MNRAS.518L..45C,2022arXiv221015699W,2022arXiv221202491L,2022arXiv221015639R}, and CEERS sources \citep{2023arXiv230315431A,2023arXiv230405378A}, all in green. The top panel shows the histogram of redshifts for these samples of high-redshift galaxy candidates. The right panel shows the magnitude distribution of the different datasets mentioned before, with the three 2022 surveys plotted with  transparency. Our deeper data (peaking around 2 magnitudes fainter than previous surveys) probe the 29 to 31~mag regime, providing candidates at the previously unexplored faint-end of the luminosity functions.}
\end{figure*}

We then selected the galaxies whose most probable (as found by integrating the zPDF) and peak (that providing the smallest $\chi^2$ value) redshifts were above $z>8$ for any of the 3 codes (1942 sources). We further cut the sample to keep only those galaxies with a $>50$\% cumulative probability of lying at $z>8$ (1831 sources). The final cut was in $\chi^2$ values, ensuring that the difference between the minimum value of the zPDF at $z<8$ and $z>8$, $\Delta\log\chi^2$ was larger than 0.4~dex (as in \citealt{2023ApJ...946L..13F}, but not as large as in \citealt{2022arXiv220801612H}), leaving 160 sources. The additional cut used in \citet{2023ApJ...946L..13F} based on the probability in $\Delta z$ intervals being higher at $z>8$ than in any other unity-redshift interval did not add anything to our method. All these cuts were based on our fiducial \texttt{eazy} run.

We then inspected each of the 160 sources visually in the NIRCam, WFC3 and ACS bands, as well as the SED fits from the different codes. We vetted galaxies which were not affected by artifacts such as (surviving) spikes and/or contamination by nearby objects. We only kept in our final sample the sources for which at least two out of the three photometric redshift codes and three of the five photometric apertures provided a peak redshift above $z>8$.

We estimated the fraction of low redshift interlopers and completeness of our selection method by simulating galaxies with the SEDs provided by the JAGUAR mock catalog \citep{2018ApJS..236...33W}. We added noise to the catalog based on the depths of our survey and applied  the selection technique (based on aggregated probabilities, goodness of fits at low and high redshift, and photo-z estimation with 3 codes) to the resulting list of sources. This analysis provided a completeness of 80\% down to F277W magnitude 31, and a contamination of 3\%, all interlopers and missed sources presenting magnitudes fainter than 30.5 mag. No correction from interlopers is applied in the luminosity function calculations presented in the following section.

The final sample of $z>8$ galaxy candidates consists of 44 objects. We show examples in Figures~\ref{fig:stamps} and \ref{fig:stamps_faint}, four dropout sources in the F115W and F150W filters, with the first figure centered on brighter galaxies ($\mathrm{F277W}<30$~mag) and the second one showing fainter examples ($\mathrm{F277W}>30$~mag).

We compared our selected $z>8$ candidate galaxies with those reported in \citet{2023arXiv230204270A} for the common area with the NGDEEP survey \citep{2023arXiv230205466B}. Only 2 sources in \citet{2023arXiv230204270A} lie within our field-of-view, another  3 (2) are located in the edges (gap between detectors) of our imaging data and only covered by 2 (0) bands (i.e., they could not arrive to our selection in any case). Of the 2 sources in common, our MDS011049 ($z\sim9.5$ in this letter) is NGD-z11a ($z\sim11.1$ for \citet{2023arXiv230204270A}). We assign it with a lower redshift mostly because it is well detected in F150W (although near our image edge). The other common source is NGD-z11b, MDS007481 in our parent catalog; it did not enter our final selection since we detected two redshift solutions, one at $z\sim6.5\pm0.2$ and another one at $z=11.1\pm0.5$,  the goodness of fit difference between them lied below our cut, and the low-z solution was favored by the photometry measured in some of our aperture.

Figure~\ref{fig:z-mag} presents the distribution of F277W (Kron-aperture) magnitudes and photometric redshifts of the final sample, compared to general catalogs and other similar compilations of $z>8$ candidates and confirmed galaxies in the first surveys carried out by {\em JWST} in 2022. The figure shows that our MDS-NIRCam-par data reach 2 magnitudes fainter than the surveys of SMACS0723, GLASS, and CEERS, with a histogram peaking at $\mathrm{F277W}\sim30$~mag, compared to 28~mag for the first datasets. The 2022 shallow surveys add up $\sim8$ times the area of our observations, but we increase by almost a factor of 10 the number of detected sources per unit magnitude bin down to 31~mag. 

Our deeper data allow us to find $z>8$ galaxy candidates at fainter magnitudes, probing lower luminosities than previous works (see Section~\ref{sec:lf}). Our sample, plotted in red in Figure~\ref{fig:z-mag}, is compared with other photometrically selected candidates and the four first spectroscopic confirmations at $z>10$ provided by {\em JWST} \citep{2022arXiv221204568C,2023arXiv230315431A,2023arXiv230405378A} as well as faint lensed galaxies above $z\sim7$ confirmed with NIRSpec spectroscopy \citep{2023MNRAS.518L..45C,2022arXiv221015699W,2022arXiv221202491L,2022arXiv221015639R}. We have seven candidates in the magnitude regime probed in the 2022 shallow surveys due to our limited area, with the bulk of our sample concentrating around 30~mag (median and quartiles $\mathrm{F277W}=30.2^{30.7}_{29.8}$~mag). Our sample doubles the number of $z\sim9$ galaxy candidates and covers an unexplored magnitude regime up to $z\sim13$. 

\begin{figure*}[t]
\centering
\includegraphics[clip, trim=2.7cm 0.4cm 2.1cm 1.0cm,scale=0.58]{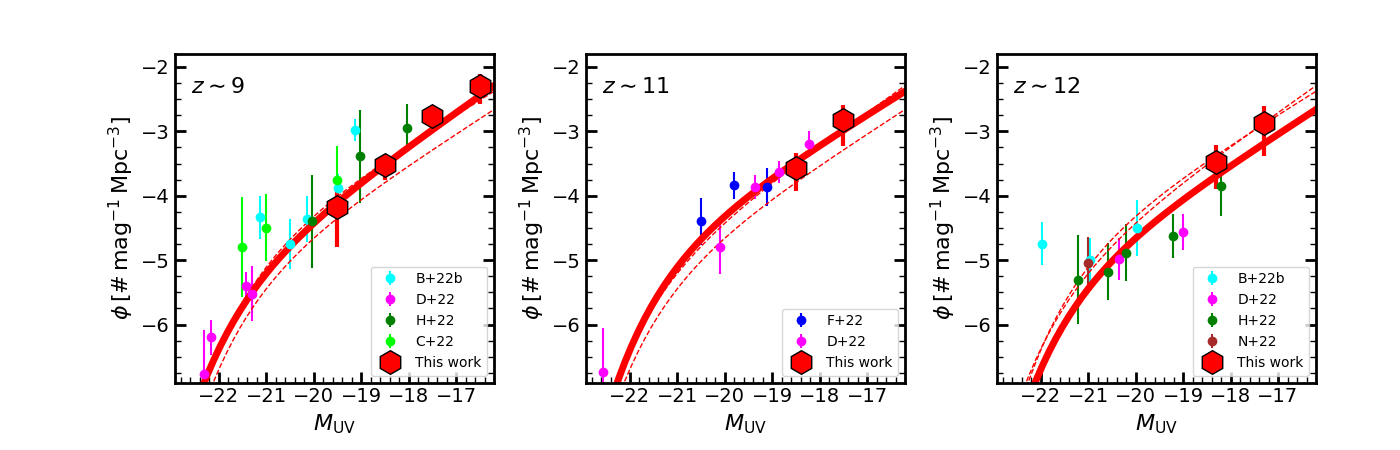}
\caption{\label{fig:lf}From left to right, UV luminosity functions at $z\sim9$, $z\sim11$ and $z\sim12$. We plot our results with red hexagons (uncertainties smaller than data points in some cases, see Table~\ref{tab:lf}), and compare with literature estimations from \citet{2023MNRAS.518.6011D,2022arXiv220709434N,2023ApJ...946L..13F,2022arXiv221206683B},  \citet{2022arXiv220801612H}, and \citet{2022arXiv221206666C}. We fit all points (ours and those in the literature) to a  \citet{1976ApJ...203..297S} function, which is plotted with red continuous lines in each panel, with the fits for other redshifts shown with dashed lines.}
\end{figure*}

We note that our data are well suited to get $z\sim9$ galaxy candidates with well covered SEDs to identify the break and obtain photometric redshift with relatively small (formal) uncertainties. However, at $z\gtrsim13$ we cannot constrain the position of the break since we do not have the F200W filter. This translates also to a higher uncertainty in the photo-z estimates at $z\gtrsim10$, when the breaks start to enter the F150W filter and we only have the F277W$-$F356W color to constrain the redshift further. This is the explanation for the larger error bars at those redshifts (especially for the $z\sim19$ candidate, shown in Figure~\ref{fig:stamps_faint}) and the gap at $z\sim10$, even when the galaxies are not significantly fainter. This was tested and confirmed with JAGUAR simulations similar to those mentioned in Section~\ref{sec:selection}. We compared results using our NIRCam filter set with those achieved when adding F200W fluxes (with depths similar to the F115W filter). The inclusion of the F200W filter decreased the typical photometric redshift errors by a factor of $>3$ and also provided larger completeness levels, reaching values beyond 90\% (compared to 80\% without that filter); the interloper fraction decreased from 3\% to 1\% when adding F200W to the simulations.

\section{Results}
\label{sec:lf}

\subsection{UV luminosity functions}

We divided the sample in 3 redshift bins, namely, $8<z<10$, $10<z<11.5$, and $11.5<z<13$, which correspond to Universe ages $\sim540$~Myr, $\sim420$~Myr, and $\sim350$~Myr. We calculated absolute UV magnitudes ($M_\mathrm{UV}$, averaged in a 0.01~$\mu$m window around 0.15~$\mu$m) from SED fitting,  obtaining uncertainties by repeating the stellar population modeling after varying the photometry according to their errors and considering the redshift probability distribution functions.

The luminosity functions were constructed using a $V_\mathrm{max}$ formalism and the stepwise maximum likelihood method \citep{1988MNRAS.232..431E}. Completeness was estimated by inserting artificial sources extracted from the real data in the detection image, covering the magnitude range between 26 and 32~mag, and then repeating the multi-layer source detection presented in Section~\ref{sec:photometry}. We find that our catalog is 80\% complete at $\mathrm{F277W}=30.4$~mag, 50\% at 30.6~mag and 10\% at 31.0~mag.

Using the implementation to determine cosmic variance effects presented by \citet{2020MNRAS.499.2401T}, consistent with the predictions from \citet{2020MNRAS.496..754B} and \citet{2021MNRAS.506..202U}, we estimate a $\sim$25\% (18\%) uncertainty in our luminosity function estimation at the bright (faint) end for $z\sim9$, the values increasing to 34\% (28\%) at the highest redshifts. These uncertainties, similar to those presented in \citet{2008ApJ...676..767T}, are included in the following figures and discussion.


The luminosity functions are presented in Figure~\ref{fig:lf} for the three redshift bins mentioned above, comparing with previous estimates in the literature. Our results are given in Table~\ref{tab:lf}. The typical systematic offsets and scatter of previous calculations are as high as 0.4~dex, with only a minor degradation as we move to higher redshifts, and some larger differences (0.6~dex level) at the bright end, $M_\mathrm{UV}<-21$~mag, at $z\sim9$ and $z\sim12$.

We fit all these estimates (our data and other data points) with \citet{1976ApJ...203..297S} functions without any prior using an MCMC method. The results are given in Table~\ref{tab:schechter}. 

We note that several of the luminosity function data points are based on the same datasets, although not on the same sources and the methodologies to select candidates are different. This might affect the uncertainties in the fits. 

To test this possibility, we repeated the fits at $z\sim9$ only including our data points and those in \citet{2023MNRAS.518.6011D} and \citet{2022arXiv220801612H}, which cover all the public JWST fields (including Stephan's Quintet, not present in \citealt{2022arXiv221206683B}) as well as the COSMOS field for the bright end. We found completely consistent results and slightly larger uncertainties: $(\alpha,M^*,\phi^*)=(-2.30^{+0.16}_{-0.24},-21.00^{+ 0.34}_{-0.45},1.8 ^{+1.1}_{-1.7})$ (normalization in units of $10^{-5}$~Mpc$^{-3}$~mag$^{-1}$) for the whole dataset, compared to $(-2.36^{+0.17 }_{-0.25},-20.96 ^{+0.40}_{-0.64},2.0 ^{+1.2}_{-2.3})$ for the limited set.

For $z\sim11$, we  obtained best-fitting values $(-2.14^{+0.24}_{-0.38},-20.74 ^{+ 0.57}_{-0.54},3.9^{+4.0}_{-3.1})$ for all datapoints (same units as above), and $(-2.39 ^{+0.16}_{-0.22},-20.24^{+ 0.45 }_{-0.29},5.3 ^{+4.1}_{-2.5})$ only using  \citet{2023MNRAS.518.6011D} and our datapoints.

Finally, for $z\sim12$, we obtained $(-2.19^{+ 0.26}_{-0.39},-20.81 ^{+0.77}_{-0.67},1.6^{+1.6 }_{-0.9})$ using all estimations (same units as above), and $(-2.05 ^{+0.40}_{-0.46},-20.25^{+ 0.95}_{-0.82},2.2^{+2.1}_{-1.3})$ when only using \citet{2022arXiv220801612H} and our datapoints.

Overall, we find little evolution between $z\sim13$ and $z\sim8$, the fits run roughly in parallel. The faint-end slope is consistent across all redshifts down to $M_\mathrm{UV}\sim-17$~mag, with a value $\alpha=-2.2\pm0.1$ and no indications of a steepening (in contrast with previous findings, e.g., \citealt{2022arXiv221206683B}).  We also obtain very similar knee absolute magnitudes, consistent within uncertainties, also considering the degeneracy with the other two Schechter parameters. The main difference between the three redshift bins is a small average number density evolution, increasing by 0.2-0.3~dex from $z\sim13$ to $z\sim8$ (between the last and first bins considered in this Letter), i.e., in $\sim200$~Myr, although this difference is of the same order as the current uncertainties for individual luminosity function estimations.


\begin{deluxetable*}{crcrcr}
\tablehead{\colhead{$\mathrm{M}_\mathrm{UV}$} & \colhead{$\log\phi$} & \colhead{$\mathrm{M}_\mathrm{UV}$} & \colhead{$\log\phi$} & \colhead{$\mathrm{M}_\mathrm{UV}$} & \colhead{$\log\phi$}\\
\colhead{AB mag} & \colhead{Mpc$^{-3}$~mag$^{-1}$} & \colhead{AB mag} & \colhead{Mpc$^{-3}$~mag$^{-1}$} & \colhead{AB mag} & \colhead{Mpc$^{-3}$~mag$^{-1}$}}
\startdata
\multicolumn{2}{c}{$8<z<10$} & \multicolumn{2}{c}{$10.0<z<11.5$} & \multicolumn{2}{c}{$11.5<z<13.0$}\\
$-19.5$ & $-4.17^{+0.26}_{-0.61}$ & $-18.5$ & $-3.57^{+0.21}_{-0.36}$ & $-18.3$ & $-3.47^{+0.26}_{-0.43}$ \\
$-18.5$ & $-3.53^{+0.15}_{-0.23}$ & $-17.5$ & $-2.82^{+0.21}_{-0.40}$ & $-17.3$ & $-2.87^{+0.28}_{-0.53}$ \\
$-17.5$ & $-2.77^{+0.11}_{-0.17}$ &       &                         &       &                                       \\
$-16.5$ & $-2.30^{+0.16}_{-0.27}$ &       &                         &       &                                       \\
\enddata
\caption{\label{tab:lf}Luminosity function data points obtained in this work.}
\end{deluxetable*}

\begin{deluxetable*}{lccc}
\tablehead{\colhead{Parameter} & \colhead{$8<z<10$} & \colhead{$10.0<z<11.5$} & \colhead{$11.5<z<13.0$}}
\startdata
$\alpha$         & $-2.30 ^{+ 0.16 }_{ -0.24 }$  & $-2.14 ^{+ 0.24 }_{ -0.38 }$ & $-2.19 ^{+ 0.26 }_{ -0.39 }$\\
$M^*$ [AB mag]   & $-21.00 ^{+ 0.34 }_{ -0.45 }$ & $-20.74 ^{+ 0.57 }_{ -0.54 }$ & $-20.81 ^{+ 0.77 }_{ -0.67 }$\\
$\phi^*$ [$10^{-5}$~Mpc$^{-3}$~mag$^{-1}$] & $1.8 ^{+ 1.1 }_{ -1.7 }$      & $3.9 ^{+ 4.0 }_{ -3.1 }$ & $1.6 ^{+ 1.6 }_{ -0.9 }$\\
$\log\rho_\mathrm{UV}$ [10$^{25}$~erg\,s$^{-1}$\,Hz$^{-1}$\,Mpc$^{-3}$] & $1.51^{+1.56}_{-0.39}$ & $1.34^{+0.45}_{-0.29}$ & $0.50^{+0.21}_{-0.18}$\\
\enddata
\caption{\label{tab:schechter}Results for the \citet{1976ApJ...203..297S} parametrization fits to the luminosity functions presented in Fig.~\ref{fig:lf}. The last row shows the integrated luminosity for absolute magnitudes $M_\mathrm{UV}<-17$~mag.}
\end{deluxetable*}

\subsection{The cosmic UV luminosity and SFR density}
\label{sec:lilly-madau}

\begin{figure*}[t]
\centering
\includegraphics[clip, trim=0.5cm 0.0cm 0.1cm 0.5cm,scale=0.5]{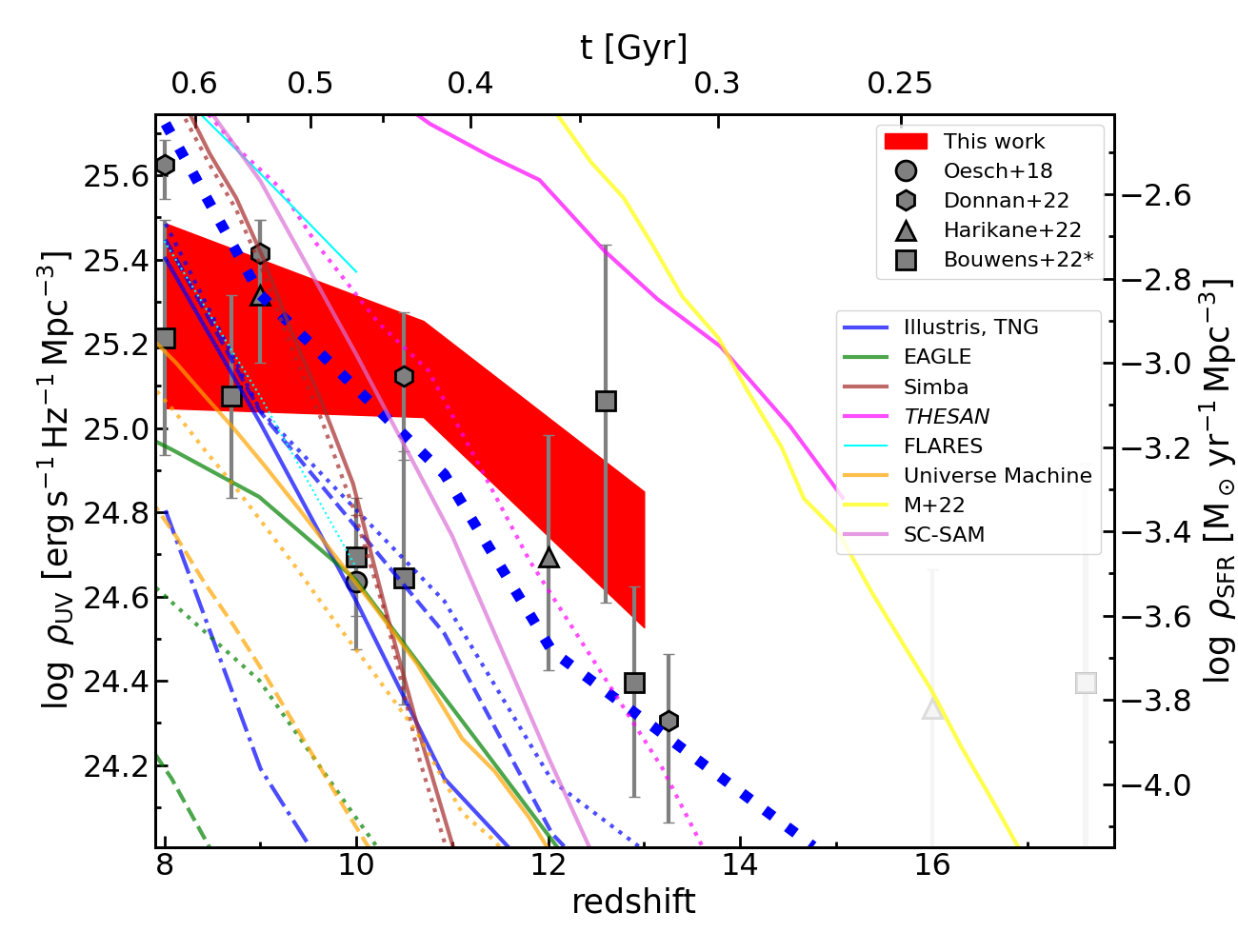}
\caption{\label{fig:density}Evolution of the UV luminosity density, transformed into SFR density on the right vertical axis (assuming a \citealt{2003PASP..115..763C} IMF). Our results are plotted with a red shaded region. Estimates from the literature are plotted with gray symbols \citep{2018ApJ...855..105O,2022arXiv221206683B,2022arXiv221102607B,2023MNRAS.518.6011D,2022arXiv220801612H}. The densities at $z>16$ are shown in light gray since they were mostly based on a $z\sim5$ redshift interloper \citep{2023arXiv230315431A,2023arXiv230406658H}. We also plot predictions from several galaxy formation simulations, listed in the following lines. The Illustris-1, {\sc TNG50, TNG100, TNG300} results are plotted with blue continuous, dashed, dotted, and dash-dotted lines, respectively, all  referring to SFRs measured in 0.5\arcsec\, diameter apertures for $\mathrm{M\!_\star}>10^{7}$~M$_\odot$ galaxies, except for {\sc TNG300}, cut at $\mathrm{M\!_\star}>10^{8}$~M$_\odot$). The thick dotted line refers to {\sc TNG100} results using whole galaxy measurements \citealt{2014MNRAS.444.1518V, 2015A&C....13...12N, 2018MNRAS.475..676S, 2018MNRAS.477.1206N, 2018MNRAS.480.5113M, 2018MNRAS.475..624N, 2018MNRAS.475..648P, 2019MNRAS.490.3196P,2019MNRAS.490.3234N}). The {\sc EAGLE} simulations \citep{2015MNRAS.446..521S,2015MNRAS.450.1937C} are plotted in green for galaxy samples cut in stellar masses $\mathrm{M\!_\star}>10^{6,7,8}$~M$_\odot$ in continuous, dotted, dashed lines. The {\sc simba} predictions are shown with a brown continuous line for all galaxies, dashed line for $\mathrm{F277W}<31$~mag sources \citep{2019MNRAS.486.2827D}. The {\sc thesan-1} and {\sc thesan-sdao-2} simulations \citep{2022MNRAS.511.4005K} are shown in magenta with continuous and dotted lines, respectively (see text for differences). The {\sc FLARES} predictions are shown in cyan \citep{2021MNRAS.500.2127L,2021MNRAS.501.3289V}. The  UniverseMachine semi-empirical model \citep{2020MNRAS.499.5702B} is shown in orange with the same mass cuts and line styles mentioned for {\sc EAGLE}. Finally, the models in \citet{2022arXiv221103620M} are shown in yellow and the predictions from the Santa Cruz semi-analytic models \citep{2019MNRAS.483.2983Y,2020MNRAS.496.4574Y} are depicted in orchid.}
\end{figure*}

We integrated the luminosity functions presented in Figure~\ref{fig:lf} down to the same absolute magnitude for all redshift bins, $M_\mathrm{UV}\sim-17$~mag, obtaining the (directly observed) UV luminosity densities at $8<z<13$. In Figure~\ref{fig:density}, we compare our results with estimates from the literature. This figure includes  pre-{\em JWST} era measurements \citep{2018ApJ...855..105O}, as well as more recent estimates obtained with {\em JWST} data \citep{2022arXiv221206683B,2023MNRAS.518.6011D,2022arXiv220801612H}. 

Figure~\ref{fig:density} also shows theoretical predictions from a range of state-of-the-art models in the literature covering a variety of approaches. The simulation compilation includes the UniverseMachine semi-empirical model \citep{2020MNRAS.499.5702B}, the Santa Cruz semi-analytical models \citep{2019MNRAS.483.2983Y,2020MNRAS.496.4574Y}, and the fiducial CDM model in \citet{2022arXiv221103620M}. We also compare with the hydrodynamical simulations Illustris/{\sc TNG} \citep{2014MNRAS.444.1518V, 2015A&C....13...12N, 2018MNRAS.475..676S, 2018MNRAS.477.1206N, 2018MNRAS.480.5113M, 2018MNRAS.475..624N, 2018MNRAS.475..648P, 2019MNRAS.490.3196P,2019MNRAS.490.3234N}, {\sc EAGLE} \citep{2015MNRAS.446..521S,2015MNRAS.450.1937C}, and {\sc SIMBA} \citep{2019MNRAS.486.2827D}. We also compare with predictions from re-simulations of multiple regions, such as {\sc THESAN} (two realizations are shown, {\sc THESAN-1} and {\sc THESAN-sdao-2}, both based on Illustris TNG galaxy formation model; \citealt{2022MNRAS.511.4005K}) and the First Light And Reionisation Epoch Simulations  ({\sc FLARES}, which adopt the AGNdT9 variant of {\sc EAGLE}; \citealt{2021MNRAS.500.2127L,2021MNRAS.501.3289V}). All these simulations probe a large box size, spanning from 50~cMpc in TNG50 to 250~cMpc in  UniverseMachine and 500~cMpc for {\sc FLARES}. These models have been validated to match observations at lower redshifts than  the epochs probed in this Letter.
 
 For this comparison with simulations, we directly took UV luminosity functions from the mentioned papers or added the SFRs (typically averaged over 100~Myr periods) of all or certain subsamples (see below) of simulated galaxies in given simulation snapshot, which can be converted to a luminosity density assuming a constant star formation event following a given IMF, \citet{2003PASP..115..763C} in our case.
 
 We note that most of these simulations provide properties integrated in galaxy regions which are larger than the actual measurements from our NIRCam observations. Indeed, the median and quartile values of the Kron apertures for our (slightly resolved) sample are $0.46^{0.58}_{0.36}$~arcsec (in diameter), i.e., observations typically refer to regions of 0.5\arcsec, which translates to $\sim2$~kpc physical size (2.4~kpc for $z=8$, 1.7~kpc for $z=13$). This means that some of the vertical offsets seen between our and literature's measurements and the model predictions can be interpreted in terms of aperture corrections.
 
 We exemplify this issue for the {\sc TNG100} simulation, shown with blue large-dot lines for whole-galaxy SFRs. Galaxies in the {\sc  TNG100} simulation at these redshifts present typical half-mass radii around 3~kpc, 6~kpc diameter, 3 times the typical photometric aperture sizes we use in this work. In Figure~\ref{fig:density}, we also show the predictions for 0.5\arcsec\, apertures (small dots). While the former curve (large dots, whole galaxy) runs closer to the observational data points, the latter (small dots, limited aperture), which is more directly comparable to the data, is consistent with observations at $z\sim8$ but predicts 0.7~dex smaller luminosity densities at $z\sim13$. This means that the star formation is occurring in more compact regions than what these simulations predict. The importance of galaxy size and morphology in {\em JWST} surveys is indeed paramount to understanding galaxy evolution in the early Universe \citep{2022arXiv220800007C}.

 Another point to consider is the mass range covered by observations and simulations. The typical stellar masses (assuming standard IMF and SFHs recipes) for our sample are $\log(\mathrm{M}\!_\star/\mathrm{M}_\odot)=7.8^{8.6}_{7.2}$ (median and quartiles, P\'erez-Gonz\'alez et al., in prep.). For some models, such as {\sc EAGLE} (in green), UniverseMachine (in orange) and Illustris/{\sc TNG} (comparing {\sc TNG300} vs other resolutions), we are able to distinguish between masses. These 3 sets of models show differences of 0.2-0.6~dex when considering a cut in $\mathrm{M}\!_\star>10^6$~M$_\odot$ instead of $\mathrm{M}\!_\star>10^7$~M$_\odot$, and they predict 0.4-0.8~dex smaller densities when cutting at $\mathrm{M}\!_\star>10^8$~M$_\odot$, these curves being more comparable to our sample.

Disregarding the difficulty in comparing with models due to aperture and mass effects (not always traceable, since some simulations do not count with the adequate resolution and/or information), most simulations predict an increase of luminosity density of almost two orders of magnitude in the 300~Myr from $z\sim13$ to $z\sim8$. The observations differ from this increase rate. We note that our {\em JWST}-based results at these redshifts present a striking agreement considering the different approaches used to identify high-redshift galaxies. All these estimates together, jointly with the first results of the UV luminosity density beyond $z=13$ shown in Figure~\ref{fig:density}, suggest a shallower increase or a more constant behaviour of the earliest phases of galaxy formation from $z=18$ until $z\sim9$, followed by a more pronounced increase of star formation activity at $z\lesssim9$. Quantitatively, the luminosity density evolution follows a $(1+z)^{-4.5\pm1.0}$ law  according to the data at $8<z<14$, $(1+z)^{-2.8\pm0.9}$ at $8<z<18$, while most models predict steeper evolution, with an average slope around $-10$, ranging from $-20$ for {\sc simba} or $-15$ for {\sc eagle} to 
$-8$ for Illustris/{\sc TNG}.

Overall, most simulations present tension with these first measurements provided by {\em JWST}, as they typically underpredict the abundance of high-redshift luminous galaxies and/or their SFHs, stellar masses, and compactness. Among the simulations, the results of {\sc THESAN} seem in agreement with observed log($\rho_{\rm SFR}$) values in the redshift range $10<z<13$ (although with no information about the spatial extension of the stellar cores we are probing with {\em JWST}), but with faster evolution (compare the steeper magenta line compared to red region). In this case, however, there is a significant difference depending on the reionization histories. In {\sc thesan-1} (continuous magenta line), the duration of reionization is short, where low mass halos have a significant contribution. In contrast, the duration is longer in {\sc THESAN-SDAO-2} (dotted magenta line, closer to the data), since it assumes non-standard dark matter models, with high mass halos being the primary drivers of reionization. Furthermore, at $z>8$ the total SFR density in {\sc THESAN-1} is dominated by the lowest mass halos ($10^8 < \mathrm{M}\!_{\rm halo}/M_{\odot} < 10^9$), which are usually unresolved in medium-resolution simulations, thus reducing the SFR density by almost an order of magnitude. 

From the observational perspective, the lack of spectroscopic confirmation (still quite demanding/unfeasible for a statistically significant sample), still hampers our physical interpretation of the early stage in galaxy formation (see \citealt{2023arXiv230406658H}). The spectroscopy is not only needed to confirm candidates but also to train our photometric redshift algorithms, necessary to probe the faint end of the luminosity function and to improve statistics and control cosmic variance effects. On the theoretical side, various possibilities are on the table, like a variable (temperature-dependant) stellar initial mass function \citep{2022ApJ...931...58S,2022ApJ...934...22S}, early dark energy models \citep{2022PhRvD.106d3526S}, dark matter properties \citep{2017MNRAS.472.4414D}, or star formation efficiencies $\sim15-30\%$ higher than expected \citep{2022ApJ...938L..10I}.





\section{Summary and conclusions}
\label{sec:conclusions}

We present a sample of 44 $z>8$ galaxy candidates identified in the NIRCam parallel observations of the MIRI Deep Survey (centered in the Hubble Ultra Deep Field and parallel pointing 2, respectively). These data reach around 2 magnitudes deeper than previous surveys carried out during Early Release Observations, Early Release Science, and Cycle 1 programs (e.g., the SMACS0723 ERO survey, CEERS, and GLASS), reaching $5\sigma$ depths around 30.8~mag. The median and quartiles of the magnitude distribution of our sample is $\langle F277W\rangle=30.2^{30.7}_{29.8}$~mag.

Aiming to reduce the effects of photometric uncertainties and {\it a priori} assumptions in the determination of redshifts, the selection of our $z>8$ galaxy candidates is based on the analysis of spectral energy distributions measured in a variety of apertures and with different photometric redshift codes. We also apply restringent cuts in the goodness of fit and the relevance of high- compared to low-redshift solutions, following standard procedures in the literature. 

The sample probes the absolute magnitude regime  $-19.5<M_\mathrm{UV}<-16.5$~mag of the luminosity function at $8<z<13$, corresponding to Universe ages between 350 and 540~Myr. We constrain the faint end, obtaining a constant slope of $-2.2\pm0.1$ in the full redshift range.  Jointly with previous results based on shallower larger-area surveys probing the bright-end of the luminosity function, we find that the regime we explore  accounts for nearly 50\% of the total luminosity density (integrated through all magnitudes brighter than $-17$~mag). 

Our estimates of the integrated ultraviolet luminosity density, which is a good proxy for the cosmic star formation rate density (via assumptions in relevant star formation properties such as the initial mass function, the metallicity, binary star fraction, or the star formation history or burstiness behavior) are compared with a wide range of predictions from state-of-the-art galaxy evolution models. The two main results are: (1) we find a shallower slope in the early stages in galaxy formation at $8<z<13$, where many models predict nearly a factor of 10 less star formation activity; (2) although some models predict similar values of the cosmic star formation rate density to those measured in this paper and in recent publications based on {\em JWST} data, when taking into account the sizes  (as measured by the aperture diameters used in the photometric extractions) and the stellar masses corresponding to the apparent magnitudes (assuming typical stellar mass calculation recipes and a standard universal initial mass function), the simulations typically identify the location of star formation in larger, less massive galaxies. For the models where the sizes and masses can be constrained to compare more fairly with observations, systematic differences with our results about the cosmic ultraviolet luminosity (or SFR) density are found at the $\times4-10$ level (depending on the specific model), indicating that the observations reveal a much more active Universe in the production of photons (through star formation or nuclear activity) the first 500~Myr on $\sim2$~kpc scales, especially at $z\gtrsim11$ (i.e., first 350~Myr of the Universe).

\begin{deluxetable*}{lcccrc}
\tablehead{\colhead{ID} & \colhead{RA (J2000)} & \colhead{DEC (J2000)} & \colhead{$F277W$} & \colhead{redshift} &  \# codes $z>8$ \\
& [degrees] & [degrees] & [mag] &}
\startdata
MDS025593 & 53.23575778 & $-$27.84791165 & $30.25\pm0.13$ & $19.6^{+4.2}_{-8.1}$ &                         2\\
MDS007258 & 53.23712648 & $-$27.85429400 & $29.89\pm0.07$ & $ 9.0^{+1.6}_{-1.6}$ &                         2\\
MDS013931 & 53.23809134 & $-$27.86844857 & $30.18\pm0.14$ & $11.5^{+7.0}_{-1.8}$ &                         2\\
MDS007507 & 53.24247528 & $-$27.85509211 & $30.24\pm0.08$ & $ 9.0^{+1.6}_{-1.2}$ &                         2\\
MDS021337 & 53.24305327 & $-$27.85711165 & $30.21\pm0.12$ & $ 8.9^{+0.7}_{-3.0}$ &                         3\\
MDS006697 & 53.24608867 & $-$27.86033795 & $29.87\pm0.14$ & $10.6^{+1.6}_{-3.2}$ &                         3\\
MDS010233 & 53.24634046 & $-$27.84432908 & $27.79\pm0.05$ & $ 9.2^{+2.0}_{-2.6}$ &                         3\\
MDS009900 & 53.24674102 & $-$27.84563609 & $29.03\pm0.04$ & $ 9.3^{+0.6}_{-4.8}$ &                         3\\
MDS028153 & 53.24926367 & $-$27.84894921 & $30.60\pm0.12$ & $ 9.3^{+2.8}_{-2.2}$ &                         2\\
MDS008711 & 53.24965269 & $-$27.85225902 & $29.60\pm0.06$ & $10.3^{+3.6}_{-4.5}$ &                         3\\
MDS015081 & 53.25034986 & $-$27.87040839 & $29.95\pm0.08$ & $ 8.7^{+1.8}_{-2.1}$ &                         3\\
MDS022582 & 53.25129541 & $-$27.85826615 & $30.57\pm0.15$ & $ 9.0^{+1.8}_{-5.2}$ &                         2\\
MDS018975 & 53.25346260 & $-$27.86461612 & $29.34\pm0.09$ & $12.2^{+3.1}_{-2.6}$ &                         2\\
MDS005128 & 53.25431438 & $-$27.87161787 & $29.88\pm0.07$ & $ 9.1^{+1.9}_{-1.6}$ &                         2\\
MDS023789 & 53.25460386 & $-$27.85749411 & $29.39\pm0.07$ & $ 9.3^{+1.0}_{-0.6}$ &                         2\\
MDS028519 & 53.25468485 & $-$27.85043609 & $30.62\pm0.12$ & $ 8.3^{+0.8}_{-2.4}$ &                         3\\
MDS026779 & 53.25653543 & $-$27.85376201 & $29.59\pm0.08$ & $11.9^{+1.8}_{-3.8}$ &                         2\\
MDS014361 & 53.25797759 & $-$27.87485511 & $30.14\pm0.21$ & $ 9.0^{+3.6}_{-0.5}$ &                         2\\
MDS029633 & 53.25929715 & $-$27.85077767 & $29.52\pm0.08$ & $ 8.5^{+2.1}_{-4.1}$ &                         2\\
MDS030229 & 53.25954821 & $-$27.85017797 & $30.62\pm0.11$ & $ 9.3^{+1.0}_{-3.2}$ &                         2\\
MDS027948 & 53.25957616 & $-$27.85306371 & $29.62\pm0.12$ & $ 9.1^{+1.1}_{-1.3}$ &                         2\\
MDS020295 & 53.26008969 & $-$27.86493963 & $30.87\pm0.18$ & $ 8.8^{+2.4}_{-3.6}$ &                         2\\
MDS018332 & 53.26100195 & $-$27.86847883 & $30.33\pm0.17$ & $10.8^{+2.4}_{-4.7}$ &                         2\\
MDS017690 & 53.26517306 & $-$27.87116585 & $30.01\pm0.11$ & $ 8.3^{+1.3}_{-5.2}$ &                         2\\
MDS011049 & 53.26722949 & $-$27.84890925 & $27.66\pm0.04$ & $ 9.4^{+0.1}_{-0.2}$ &                         2\\
MDS006210 & 53.28683244 & $-$27.87802853 & $29.63\pm0.09$ & $ 9.1^{+2.3}_{-4.2}$ &                         3\\
MDS021311 & 53.28746863 & $-$27.87372630 & $31.12\pm0.22$ & $ 8.4^{+1.7}_{-1.6}$ &                         3\\
MDS022349 & 53.29153766 & $-$27.87366009 & $30.36\pm0.15$ & $ 8.8^{+1.3}_{-2.3}$ &                         3\\
MDS008116 & 53.29301264 & $-$27.87124117 & $28.91\pm0.11$ & $12.6^{+2.4}_{-2.6}$ &                         2\\
MDS004915 & 53.29351197 & $-$27.88750272 & $29.39\pm0.10$ & $10.7^{+3.0}_{-3.5}$ &                         3\\
MDS020574 & 53.29521802 & $-$27.87761048 & $30.81\pm0.18$ & $10.7^{+2.7}_{-3.5}$ &                         3\\
MDS005199 & 53.29802771 & $-$27.88750404 & $29.52\pm0.08$ & $ 8.7^{+1.5}_{-1.9}$ &                         3\\
MDS005247 & 53.29901733 & $-$27.88760795 & $29.75\pm0.08$ & $ 9.1^{+0.7}_{-1.4}$ &                         2\\
MDS006765 & 53.30019902 & $-$27.88012852 & $29.29\pm0.10$ & $ 9.0^{+1.7}_{-1.8}$ &                         3\\
MDS018069 & 53.30742315 & $-$27.88628994 & $30.78\pm0.14$ & $ 8.2^{+2.1}_{-3.5}$ &                         2\\
MDS007889 & 53.30748690 & $-$27.87763847 & $29.84\pm0.08$ & $ 8.8^{+2.7}_{-1.6}$ &                         2\\
MDS008635 & 53.30804862 & $-$27.87443982 & $29.98\pm0.13$ & $ 8.9^{+1.9}_{-4.7}$ &                         2\\
MDS008261 & 53.30820570 & $-$27.87625163 & $28.73\pm0.06$ & $ 9.2^{+2.5}_{-4.0}$ &                         3\\
MDS006079 & 53.30921148 & $-$27.88711824 & $28.91\pm0.05$ & $10.7^{+3.1}_{-3.3}$ &                         3\\
MDS025580 & 53.31027610 & $-$27.87573262 & $29.17\pm0.10$ & $10.1^{+2.6}_{-3.8}$ &                         3\\
MDS005234 & 53.31030222 & $-$27.89189564 & $29.27\pm0.08$ & $12.4^{+8.6}_{-4.0}$ &                         2\\
MDS025608 & 53.31030283 & $-$27.87569216 & $30.10\pm0.23$ & $10.7^{+3.3}_{-3.5}$ &                         2\\
MDS006529 & 53.31597314 & $-$27.88737382 & $30.01\pm0.09$ & $ 8.9^{+0.9}_{-6.0}$ &                         3\\
MDS025422 & 53.31811452 & $-$27.87887496 & $30.54\pm0.19$ & $11.3^{+4.0}_{-3.8}$ &                         2\\
\enddata
\caption{\label{tab:sample}Sample of $z>8$ galaxy candidates presented in this letter. We provide coordinates, integrated magnitudes in the $F277W$ filter and photometric redshifts estimated with {\sc eazy}. The comment gives information about how many of the 3 photometric redshift codes used in this letter agree in the $z>8$ determination.}
\end{deluxetable*}

\begin{acknowledgements}
The authors dedicate this Letter to the memory of the members of the European Consortium MIRI Team Hans Ulrik N{\o}rgaard-Nielsen and Olivier Le Fèvre, R.I.P. We thank the anonymous referee for their constructive comments that certainly helped to the robustness of this work. This work has made use of the Rainbow Cosmological Surveys Database, which is operated by the Centro de Astrobiología (CAB), CSIC-INTA, partnered with the University of California Observatories at Santa Cruz (UCO/Lick, UCSC). PGP-G and LC acknowledge support  from  Spanish  Ministerio  de  Ciencia e Innovaci\'on MCIN/AEI/10.13039/501100011033 through grant PGC2018-093499-B-I00. 
LC acknowledges financial support from Comunidad de Madrid under Atracci\'on de Talento grant 2018-T2/TIC-11612. L. Colina and Javier Álvarez-Márquez acknowledge support by grant PIB2021-
127718NB-100. GSW acknowledges funding from the UK Science and Technology Facilities Council, and the UK Space Agency.
DL and JH were supported by a VILLUM FONDEN Investigator grant to JH (project number 16599).
AAH acknowledges support from PID2021-124665NB-I00 funded by the Spanish Ministry of Science and Innovation and the State Agency of Research MCIN/AEI/10.13039/501100011033.  TG and SG acknowledge the support of the Cosmic Dawn Center of Excellence (DAWN) funded by the Danish National Research Foundation under the grant 140. TRG acknowledges support from the Carlsberg Foundation (grant no CF20-0534). KIC acknowledges funding from the Dutch Research Council (NWO) through the award of the Vici Grant
VI.C.212.036. JPP and TVT acknowledge funding from the UK Science and Technology Facilities Council, and the UK Space Agency. OI acknowledges the funding of the French Agence Nationale de la Recherche for the project iMAGE (grant ANR-22-CE31-0007) and the support of the Centre National d'Etudes Spatiales (CNES). BV thanks the Belgian Federal Science Policy Office (BELSPO) for the provision of financial support in the framework of the PRODEX Programme of the European Space Agency (ESA). ACG acknowledges support by grant PIB2021-127718NB-100. RAM acknowledges support from the ERC Advanced Grant 740246 (Cosmic\_Gas). 

Some/all of the data presented in this paper were obtained from the Mikulski Archive for Space Telescopes (MAST) at the Space Telescope Science Institute. The specific observations analyzed can be accessed via \dataset[DOI: 10.17909/je9x-d314]{https://doi.org/10.17909/je9x-d314}.
\end{acknowledgements}

\bibliography{perezgonzalez_mirigto_nircampar_highz}
\bibliographystyle{aasjournal}



\end{document}